\documentstyle[prl,aps]{revtex}
\begin{document}
\title{A Model for the Thermodynamics of Globular Proteins}
\author{Alex Hansen\footnote{Electronic Address: Alex.Hansen@phys.ntnu.no}} 
\address{Department of Physics,
Norwegian University of Science and Technology, NTNU, N--7034 Trondheim, 
Norway}
\author{Mogens H. Jensen,\footnote{Electronic Address: mhjensen@nbi.dk}
Kim Sneppen\footnote{Electronic Address: sneppen@nbi.dk} 
and Giovanni Zocchi\footnote{Electronic Address: zocchi@nbi.dk}}
\address{Niels Bohr Institute and NORDITA, Blegdamsvej 17, DK-2100 {\O}, 
Denmark}
\date{\today}
\maketitle
\begin{abstract} 
We review a statistical mechanics treatment of the stability of 
globular proteins based on a simple model Hamiltonian taking into account
protein self interactions and protein-water interactions.  The model 
contains both hot and cold folding transitions.  In addition it predicts
a critical point at a given temperature and chemical potential 
of the surrounding water.  The universality class of this critical point
is new. 
\end{abstract} 
\vskip0.5cm
Biologically relevant proteins are macromolecules \cite{ablrrw94} whose 
structures are determined by the evolutionary process \cite{n84,g85}. 
The folded conformation of globular proteins is a state of
matter peculiar in more than one respect. The density is
that of a condensed phase (solid or liquid),
and the relative positions of the atoms are, on average, fixed.
These are the characteristics of the solid state.
However, solids are either crystalline or amorphous,
and proteins are neither: the folded structure, while ordered in the sense
that each molecule of a given species is folded in the same way,
lacks the translational symmetry of a crystal.  Unlike any other
known solids, globular proteins are not really rigid, being able to perform
large conformational motions while retaining locally the same folded 
structure.  Finally, these are mesoscopic systems, consisting of a few 
thousand atoms.

Quantitatively, the peculiarities of this state of matter are
perhaps best appreciated from thermodynamics.
Delicate calorimetric measurements \cite{Priv1,Priv2,Privalov}
on the folding transition of globular proteins reveals
the following picture. The transition is first order,
at least in the case of single domain proteins.
The stability of the folded state,
i.e., the difference in Gibbs potential
$\Delta G$ between the unfolded and the folded state is
at most a fraction of $k T_{room}$ per aminoacid.
This is refered to as ``cooperativity". The 
Gibbs potential difference $\Delta G$,
as a function of temperature, is non monotonic:
it has a maximum around room temperature (where $\Delta G > 0$
and consequently the folded state is stable), then crosses zero and
becomes negative both for higher {\it and\/} lower temperatures.
Correspondingly, the protein unfolds not only at high,
but also at low temperatures.  The melting transition under cooling is
refered to as ``cold unfolding" or ``cold denaturation."
For temperatures around the cold unfolding 
transition and below, the
enthalpy difference $\Delta H$ between the unfolded and the
folded state is negative; this means that cold unfolding
proceeds with a release of heat (a negative latent heat),
as is also observed experimentally; at the higher unfolding
transition, on the contrary, $\Delta H > 0 $ which corresponds
to the usual situation of a positive latent heat.
There are two peaks in the specific heat, corresponding to the
two unfolding transitions, and a large gap $\Delta C$ in the
specific heat between the unfolded and the folded state.
This gap is again peculiar to proteins: usually,
for a melting transition $\Delta C$ is small.

It is (however not universally) 
believed that from the microscopic point of view,
the main driving force for folding is the
hydrophobic effect; in the native state of
globular proteins hydrophobic residues
are generally found in the inside of the molecule,
where they are shielded from the water, while hydrophobic residues
are typically on the surface. 
Hydrogen bonds within the regular elements of secondary
structure ($\alpha$ helices and $\beta$ sheets), while
necessary for the stability of the native state, can hardly
be thought of as providing the positive $\Delta G$ of the
folded structure, since the unfolded structure would form
just as many hydrogen bonds with the water. When the
protein unfolds, the hydrophobic residues of the interior
are exposed; this accounts for most of the gap in the
specific heat $\Delta C$ \cite{Privalov}, according to
the known effect that dissolving hydrophobic substances
in water raises the heat capacity of the solution \cite{Edsall}.
A recent discussion of hydrophobicity in protein
folding may be found in Ref.\ \cite{Tang}.

As in other branches of physics, once the thermodynamics
of a system is known it is desirable to develop a corresponding
statistical mechanics model.
In the following, we describe a recently proposed model of this kind
that accounts for the strange thermodynamical behavior described above
\cite{hjsz98a,hjsz98b,hjsz98c,hjsz99}.  Its starting point is the simple but 
appealing ``zipper model" \cite{zipper},
which was introduced to describe the helix -- coil transition.
In this model, the relevant degrees of freedom (conformational angles) are 
modeled through binary variables.  Each variable is either matching the 
ordered structure (helix), or in a ``coiled" state.  
A related parametrization for the 3-d folding transition
has been proposed by Zwanzig \cite{Z95}, describing it in terms of 
variables $\psi_i$, each of which is ``true" (1) when there is local match 
with the correct ground state, or ``false" (0) 
if there is no match.  The term ``local" 
is here defined through the parametrization index $i$.
A zipper scenario that deals with the initial pathway of protein
folding has been proposed by Dill et al.\ \cite{dfc93}.  We can parametrize
this model in the same way as done by Zwanzig by assigning the value one
to each of the binary variables 
$\psi_i$ describing closed contacts in the zipper.  
Build into the model 
is that opening and closing of contacts occur in a particular order:  They
behave as the individual locks in a zipper.  This ordering is characterized
through imposing the constraints
\begin{equation}
\label{constraint}
\psi_i \ge \psi_{i+1}\;.
\end{equation}
The variables $\psi_i$ alone 
cannot describe the degrees of freedom that become liberated when a portion
of the zipper is open.  
The open part of the zipper may move freely ($\psi_i =0$)
whereas they cannot move in the part of the zipper where
the contacts are closed ($\psi_i=1$).  
In order to take into account this effect, we 
introduce a second, independent set of variables $\xi_i$. 
For simplicity, we also make these variables binary, taking
the values 1 or -$B$.  We are now in the position to propose a Hamiltonian 
for this zipper model,
\begin{equation}
\label{psiham}
H~=~-\sum_{i=1}^N \psi_i\xi_i\;,
\end{equation}
subjected to the constraints (\ref{constraint}).  

We note that for any finite value of $B$, parts of the protein
may unfold inside the already folded region i.e. in the parts of
the zipper where $\psi_i=1$.
In order to prevent this, we assume $B$ to be sufficiently large
compared to any other energy scale in the system --- in particular $kT$,
where $T$ is the temperature --- so that the $\xi_i$ variables never
assume the value $-B$ as long as $\psi_i=1$.

We will in the following use this Hamiltonian as a starting point for 
analyzing the hot and cold denaturation transitions of proteins when 
dissolved in water \cite{hjsz98b}.  It is awkward to work with the
Hamiltonian (\ref{psiham}) directly because of the constraints 
(\ref{constraint}).  We therefore make a transformation to a 
different set of variables where the constraints (\ref{constraint}) 
are implicitly taken into account.
We define a set of binary, unconstrained
variables $\varphi_i$, by the following relation:
\begin{equation}
\label{psiphi1}
\psi_i=\varphi_1\cdots\varphi_i\;.
\end{equation}
In particular, $\psi_1=\varphi_1$.  In the limit when $B\to\infty$, 
the Hamiltonian (\ref{psiham}) becomes
\begin{equation}
\label{phiham}
H~=~ - \varphi_1 - \varphi_1\varphi_2 -
\varphi_1\varphi_2\varphi_3 - \cdots -
\varphi_1\varphi_2\cdots\varphi_N\;,
\end{equation}
where there are no additional constraints
\cite{hjsz98c}.  The role of the variables $\xi_i$ --- which is to provide
entropy to the unfolded part ($\psi_i=0$) of the zipper --- 
is now played by the degeneracy introduced into the Hamiltonian in the
following way:  When a particular $\varphi_j=0$, the Hamiltonian 
(\ref{phiham}) will be degenerate with respect to the variables
$\varphi_i$ where $i>j$.

The interactions between protein and water may be taken into account by adding
to (\ref{phiham}) a coupling parametrized through water variables 
$w_1, w_2, ... , w_N$ \cite{hjsz98b}.  Returning for a moment to the 
original variables $\psi_i$, we propose an interaction $(1-\psi_i\xi_i)w_i$.
The rationale behind this form is that when a contact is open ($\psi_i=0$),
the part of the protein parametrized by $i$ is exposed to water and interact, 
while if the
contact is closed ($\psi_i=1$), there is no access to the water and the
interaction is zero.  Returning to the new variables $\varphi_i$, the 
resulting Hamiltonian is 
\begin{equation}
\label{eq2}
H~=~- {\cal E}_0 ~
(\varphi_1+\varphi_1\varphi_2+
\varphi_1\varphi_2\varphi_3+\cdots +
\varphi_1\varphi_2\cdots\varphi_N)\\ 
+ [ (1-\varphi_1) w_1 + (1-\varphi_1\varphi_2) w_2 + ... + 
(1-\varphi_1\varphi_2\cdots\varphi_N) w_N ]\;,
\end{equation}
where we have introduced a scale parameter ${\cal E}_0$ in order to vary 
the relative strength of the protein self interactions and the protein-water 
interactions.  In order to model hydrophobicity, we 
assume the $w_i$ variables take values 
${\cal E}_{\min} + s \Delta$, $s=0,1,...,g-1$. Here, $\Delta$ is
the spacing of the energy levels of the water-protein interactions.
The equidistant energy levels reflect the experimentally observed
approximate constant heat capacity at intermediate temperatures,
whereas the finite number of levels $g$ takes into account that 
protein-water interactions vanish at high temperatures,
in practice above 120 degree celsius.

The number of terms in the Hamiltonian (\ref{eq2}), $N$, is the number of
contact in the zipper model. This number may be equal to the number of
amino acids, but is a priori unknown.
It is important to realize that if one parametrize the folding
with fewer steps $N$, each unit will be larger and energies and 
entropies appropriately increased (inversely proportional to $N$).

The calculation of the partition function is straightforward.
We parametrize the states of the system
by the number $n$ of consequtive matches
$\varphi_1 =1, \varphi_2=1,..., \; \varphi_{n}=1$
and ending with $\varphi_{n+1}=0$
and the values $\{s_{n+1},... ,s_N\}$ where each
$s_i \in \{0,1,2,..., g-1\}$ for the $(N-n)~$ $\mu$
variables coupled to the unfolded portion of the protein.
The energy of this state is
\begin{equation}
\varepsilon (n,s_{n+1},..., s_N )
~=~ - n~ {\cal E}_0  ~+~ \sum_{i = n+1}^N
({\cal E}_{min} + \Delta {\cal E}~ s_i)
\end{equation}
where we have introduced the energy scale ${\cal E}_0$ for the protein
variable in order to
make the formulas dimensionally more transparent
(up to now we used ${\cal E}_0=1$).
Denoting $\beta=1/T$ as the reciprocal temperature,
the partition function is
\begin{equation}
\label{eq4}
Z = \sum_{n=0}^{N-1} \;
2^{N-n-1} g^{n} \;
\sum_{s_{n+1}=0}^{g-1}
\sum_{s_{n+2}=0}^{g-1} \cdots \sum_{s_N=0}^{g-1}
\exp(\;- \beta \varepsilon (n,s_1,\cdots, s_N ) )
~~+~ g^N \exp( \;\beta {\cal E}_0 N)
\end{equation}
In the above equation  the factor $2^{N-n-1}$ is the
degeneracy of the unfolded protein degrees of freedom and
the factor $g^{n}$ is the degeneracy of water which is not
exposed to the inside of the protein. Factorizing the sums
over $s_i$ into partition functions $Z_w$
\begin{equation}
\label{eq5}
Z \;=\; \frac{1}{2} ~ (2 Z_w )^{N} \sum_{n=0}^{N-1}
\left( \frac{ g \exp(\beta {\cal E}_0) }{2 Z_w} \right)^{n}
~~+~ \left( g \exp( \;\beta {\cal E}_0 ) \right)^N
\end{equation}
where the phase space for a water degree of freedom exposed
to a unfolded protein degree of freedom is
\begin{equation}
Z_w \; =\; \sum_{s=0}^{g-1} \exp( -\beta ( {\cal E}_{min} +
s \Delta {\cal E} ) )
\; =\; \frac{(\exp(-\beta {\cal E}_{min}) -
\exp(-\beta {\cal E}_{max})}{
( 1 - \exp(-\beta \Delta {\cal E}))}
\end{equation}
where ${\cal E}_{max}={\cal E}_{min} + g \Delta {\cal E}$.
From Eq.\ \ref{eq5} one sees directly that the state of the system
is determined by the size of the quantity
\begin{equation}
\label{rrr}
r\;=\;\frac{ g \exp(\beta {\cal E}_0)}{2 Z_w} \;=\; \exp(\beta \Delta f)
\end{equation}
If $\Delta f > 0$ then  the system will be in the folded state
because the sum in Eq.\ (\ref{eq5}) is dominated by the last term, whereas
for $\Delta f<0$ the system will be unfolded.

The sum in Eq.\ (\ref{eq5}) can be readily performed
and the total partition function is
\begin{equation}
Z \;=\; \frac{1}{2} (2 Z_w )^{N} ~ \frac{
1 - \left(  g \exp({\cal E}_0 \beta)~ / ~(2 Z_w) \right)^{N}
}{
1 - \left(  g \exp({\cal E}_0 \beta)~ / ~(2 Z_w) \right)
}
~~+~ \left( g \exp( \;\beta {\cal E}_0 ) \right)^N
\end{equation}
The free energy is $F~=~-T \ln (Z)$,
the energy $E~=~-d \ln(Z)/d\beta$ and the heat capacity
$ C~ =~ dE/dT$.
Because there is no pressure in the model,
the energy $E$ takes the place of
the enthalpy $H = E + p V$ and the free energy
$F = E - T S$ takes the place of
the Gibbs potential $G = H - T S$.

In Fig.\ \ref{fig2} we show the heat capacity the three different
choices of ${\cal E}_{min}$, representing three different
values of the chemical potential as we discuss later.
The characteristic feature is that there are two peaks
corresponding to warm and cold unfolding, and a gap
$\Delta C$ in the heat capacity between the unfolded and the
folded form. At higher temperatures, i.e., $T > g \Delta {\cal E}$,
the gap goes to zero because the water becomes effectively
degenerate again. 
In Fig.\ \ref{fig3} we show the order parameter
$\langle n \rangle$ as function of temperature for three values of
the chemical potential.
The figure indeed confirms that the protein is folded between
the two transitions.

We now calculate explicitly the difference in the thermodynamic
functions between the unfolded and the folded state.
We consider these quantities per degree of freedom.
The thermodynamic functions associated to a folded (f) protein variable
is the energy $e_f=-{\cal E}_0$, the entropy $s_f=\ln(g)$
and the free energy $f_f=-{\cal E}_0-T\ln(g)$.
The free energy associated to an unfolded (u) protein variable
is given by the corresponding partition function of water
multiplied by the degeneracy factor of
an unfolded part of the protein: $f_u=-T\ln(Z_w ~ 2)$.
The difference in free energy between folded and unfolded
state is accordingly
\begin{equation}
\Delta f ~=~ f_u ~-~ f_f ~=~
T~ \ln( \frac{ g~ \exp( \beta {\cal E}_0)}{2 ~Z_w} )
\end{equation}
which is the quantity we earlier identified as the one
which decides whether the system cooperatively selects
the folded or the unfolded state.
To clarify the physical contents of this formula we rewrite it
for small energy level spacings $\Delta {\cal E} <<T$:
\begin{equation}
\label{deltaf}
\Delta f ~=~
{\cal E}_0 ~+~ {\cal E}_{min} ~+~
T \ln ( \frac{g \Delta {\cal E}}{2 ~ T} )
~-~T \ln \left( 1- \exp(-({\cal E}_{max} -{\cal E}_{min})/T) \right)
\end{equation}
From this expression for the difference in free energy one easily obtains the
corresponding differences in energy, entropy and specific heat. In
particular, we obtain a gap in the specific heat
between the folded and unfolded state of a protein degree of freedom
$\Delta c =  ( \Delta {\cal E} / T)^2 / (e^{\Delta {\cal E} / T} - 1)^2 ~
e^{\Delta {\cal E} / T} ~\sim~ \exp( \Delta {\cal E} / T) \sim 1$
for temperatures $T \in [\Delta {\cal E},{\cal E}_{min}+{\cal E}_{max}]$,
see Fig.\ \ref{fig2}.

To simplify the discussion let us consider the limit of large
${\cal E}_{max}$ in (\ref{deltaf}). It is easily seen that $\Delta f$
has a maximum at the temperature
$T_m \approx g \Delta {\cal E} / 2 e$ .
The corresponding value of $\Delta f$ is
$ \Delta F(T_m) \approx ~ ({\cal E}_{min} +
{\cal E}_0) ~ + ~ g \Delta {\cal E} / 2 e $ ,
so the condition for the existence of a region of
stability of the ordered structure ($\Delta f > 0$) is:

\begin{equation}
\frac{g \Delta {\cal E}}{2 e} > - ({\cal E}_{min} + {\cal E}_0) .
\end{equation}
This is of course always satisfied if
$ ({\cal E}_{min} + {\cal E}_0) > 0$ , however
the more interesting situation is
$ ({\cal E}_{min} + {\cal E}_0) < 0$, since then
$\Delta F < 0$ at sufficiently low temperature,
i.e.\ the phenomenon of cold unfolding
appears. Under these conditions $\Delta E$ is
also negative at sufficiently low temperature
which means that we have a negative
latent heat for cold unfolding. 

Coming back to the partition function (\ref{eq4}) 
and (\ref{eq5}), we may write:

\begin{equation}
{\cal E} = - N {\cal E}_0 ~+~ (N-n) ({\cal E}_0 + {\cal E}_{min})
+ \sum_{i=n + 1}^N ~ \Delta {\cal E}~ s_i ~=~ - N {\cal E}_0 ~+~
\sum_{i=n + 1}^N ~[\Delta {\cal E}~ s_i + {\cal E}_0 + {\cal E}_{min}]
\end{equation}
and
\begin{equation}
Z =  e^{\beta N {\cal E}_0}  \sum_{n=0}^{N-1} ~
2^{N-n-1} g^{n} \;
\sum_{\lbrace s_i \rbrace} ~
e^{- \beta \sum_{i=n + 1}^N ~ ({\cal E}_i - \mu )} ~
~~+~ g^N \exp( \;\beta {\cal E}_0 N)
\end{equation}
where we have set
${\cal E}_i = \Delta {\cal E} ~ s_i ~,
~ \mu = - ( {\cal E}_0 + {\cal E}_{min})$.
From this expression for $Z$ we can identify $\mu$ with the
chemical potential \it{ of the water } \rm , or, to be more
precise, the difference in chemical potential of the water
when it is in contact with the hydrophobic
interior of the protein and when it is not.
Therefore, $\mu > 0$ is the physically
relevant situation. 
Experimentally, $\mu$ can be changed by adding denaturants,
changing pH, etc., which indeed alters the stability of the 
ordered structure.

For an intermediate value of the chemical potential,
$r$ --- defined in Eq.\ (\ref{rrr}) --- 
just touches the line $r=1$, that is $dr/dT=0$ when $r=1$,
corresponding to a merging of two first order transitions.
This defines a critical point.  Around this point, $r$ varies quadratically in
$T-T_c$ and linearly in $\mu-\mu_c$, as seen from expanding Eq.\ (\ref{rrr}).
In experiments of protein folding this point is 
accessible by changing the pH value of the solution.
In fact, Privalov's data on low pH values indeed 
indicate that such a critical point exists.
The scaling properties around this point
thus opens for a possibility to gain insight into the 
nature of the folding process, in particular whether 
the pathway scheme we suggest can be falsified.

In Fig.\ \ref{fig2} we show heat capacity as a function of temperature
for chemical potential at the critical value $\mu=\mu_c$. 
For the chosen values of ${\cal E}_0=1$ and level density
$\Delta=0.02$ and $g=350$
the critical point is situated at $T_c=1.33303\dots$, $\mu_c=1.2838\dots$.
That is, it is situated at a {\it minimum\/} of the heat capacity curve.
This is at first sight surprising, usually heat capacity has 
a pronounced increase at the critical point.
The minimum reflects a partial ordering, 
as envisioned in Fig.\ \ref{fig3} where we show the degree 
of folding, counted by the average number of folded 
variables $\varphi_i=1$, $i=1,...,n$ from $i=1$ 
until the first variable $i=n+1$ which takes value $\varphi_{n+1}=0$.
The average value of this $\langle n \rangle$ is
$N/2$ at the critical point, reflecting 
that the system is on average half ordered at this point.
Correspondingly the heat capacity dips to a value
in between the value of an unfolded and a
completely folded state.

To characterize the functional form of the dip
in the heat capacity, we investigate analytically
$C_{sing}(T)=C(T,\mu)-C(T,\mu_c)$ with $\mu>>\mu_c$
for different values of the size $N$.
For finite $N$ we may express the singular
part of the heat capacity in the form:
\begin{equation}
C_{sing} ~ =~ |T_c-T|^{-\alpha} ~ g \left( (T_c-T) N^{1/\nu} \right)
\end{equation}
where $g(x) \rightarrow const$ when $x\rightarrow \infty$
and $g(x)\propto x^{\alpha}$ when $x\rightarrow 0$.
We find analytically $\alpha=\nu=2$ from differentiating 
the partition function (\ref{eq5}).
Fig.\ \ref{fig6} demonstrate this finite size scaling.
Similarly we show in Fig.\ \ref{fig7} the behavior
of the order parameter $\langle n \rangle$
as function of $T-T_c$ and $N$:
\begin{equation}
\langle n \rangle ~ =~ |T-T_c|^{\beta} ~ 
f \left( (T-T_c) N^{1/\nu} \right) 
\end{equation}
with $f(x) \rightarrow const$ when $x\rightarrow \infty$
and $f(x)\propto x^{-\beta}$ when $x\rightarrow 0$
where exponents $\beta=-2$, also found analytically.  
It may be surprising that $\beta$ 
is negative, but this reflect in part
the unusual use of an extensive (in $N$) order parameter,
in part that for $\mu=\mu_c$ then the order parameter only obtains 
a non-zero value at $T=T_c$ when $N\rightarrow \infty$.

Likewise, we find that the susceptibility 
$\chi=d\langle n\rangle/d\mu$ scales as 
$|T-T_c|^{-\gamma}$ where $\gamma=4$ and 
that $\langle n \rangle \propto (\mu-\mu_c)^{1/\delta}$ 
for $\mu>\mu_c$ where $\delta=-1$. Thus the usual exponent relations,
$\alpha+2 \beta + \gamma=2$, $\alpha+\beta (\delta+1) =2$,
and $\gamma (\delta + 1) =(2-\alpha)(\delta-1)$ 
are fulfilled \cite{s71}.
However the hyperscaling relation $d \nu = 2-\alpha$, where $d$ is the
dimensionality of the system, is not fulfilled.  However, this relation has
no meaning, as there are no spatial degrees of freedom.

In terms of experiments on proteins, the relevant
scaling behaviour is the how the degree of folding
(order parameter) and the heat capacity behaves
as function of temperature, when one changes chemical potential
away from its critical value.
The qualitative prediction is that the width 
of the singular part of the heat capacity has a minimum
at the critical value $\mu=\mu_c$.
The broadening of the heat capacity is
\begin{equation}
C_{sing} (T-T_c)^2 ~=~ 
h\left(\frac{T-T_c}{\Delta \mu^{1/2}}\right) ~~ for~~ 
\mu> \mu_c~~ 
\end{equation}
where $h(x)\propto x^{-2}$ for $x \rightarrow \infty$ and 
$h(x)=const$ for $x \rightarrow 0$
and where $\Delta \mu=\max(\mu-\mu_c, \Delta \mu_{\min})$
with $\Delta \mu_{\min} \propto 1/N$ takes into account the
finite size sensitivity of the scaling. We show in Fig.\ \ref{fig8},
an example of such a data collapse.
These predictions are experimentally accessible through the use of standard
calorimetric techniques, where one should seek to 
obtain a data collapse above the critical point, 
i.e. the point of minimal width.
The heat capacity below the critical $\mu$ is complicated by
the merging of two first order transitions.
However, the distance between these moves away from each other
in $T$ as $\Delta \mu^{1/2}$.

Likewise, we expect the degree of folding $\langle n \rangle$
to show data collapse of the form
\begin{equation}
\langle n \rangle (T-T_c)^2
\mu> \mu_c~~ 
\end{equation}
where $k(x)$ behaves asymptotically as $h$. We show this in
Fig.\ \ref{fig9}.
This quantity can be observed 
experimentally through fluorescence measurements.

\begin{figure}
\caption{Heat capacity, $C$, as a function of $T$ for three values of
the chemical potential $\mu$.
Here $g=350$, $\Delta=0.02$ and $N=100$. The value $N=100$ has been chosen
as to be close to realistic values for this parameter.
\label{fig2}
}
\end{figure}
\begin{figure}
\caption{
Degree of folding, $\langle n\rangle$, as a function of $T$ for three
values of the chemical potential $\mu$.  The
parameters are chosen as in Fig.\ \ref{fig2}.
\label{fig3}
}
\end{figure}
\begin{figure}
\caption{Finite size scaling of the heat capacity for $\mu=\mu_c$,
$g=350$ and $\Delta=0.02$. Here $\alpha=2$ and $\nu=2$.
\label{fig6}
}
\end{figure}
\begin{figure}
\caption{a) Finite size scaling of folding, $\langle n\rangle$for $\mu=\mu_c$,
$g=350$ and $\Delta=0.02$. Here $\beta=-2$.
\label{fig7}
}
\end{figure}
\begin{figure}
\caption{$C_{sing}(T-T_c)^2$ {\it vs.\/} $(T-T_c)/\Delta\mu^{1/2}$.
We have chosen $N=100$, $g=350$ and $\Delta=0.02$.  Note the good quality
of the data collapse in spite of smallness of the system.
\label{fig8}
}
\end{figure}
\begin{figure}
\caption{
$\langle n\rangle(T-T_c)^2$ {\it vs.\/} $(T-T_c)/\Delta\mu^{1/2}$.
We have chosen $N=100$, $g=350$ and $\Delta=0.02$.  Note the good quality
of the data collapse in spite of smallness of the system.
\label{fig9}
}
\end{figure}
\end{document}